*Original Article* 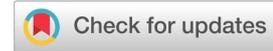

# Analysis of the Metabolic Profile and Biological Activity of Hawthorn Species Twigs: *Crataegus azarolus* and *Crataegus monogyna*

Hiwa Sheikh Ahmed Qalatobzany [a] , Kadhm Abdullah Muhammad [b], Djshwar Dhahir Lateef [c]
Kamaran Salh Rasul [d*] , Abdulrahman Smail Ibrahim [e] , Mariana Casari Parreira [e],
Weria Weisany [f]

[a] Department of Chemistry, College of Science, University of Garmian, Kalar, Sulaymaniyah, Iraq.
[b] Department of Agribusiness and Rural Development, College of Agricultural Engineering Sciences, University of Sulaimani, Sulaymaniyah, Iraq.
[c] Department of Biotechnology and Crop Sciences, College of Agricultural Engineering Sciences, University of Sulaimani, Sulaymaniyah, Iraq.
[d] Department of Horticulture, College of Agricultural Engineering Sciences, University of Sulaimani, Sulaymaniyah, Iraq.
[e] School of Agrarian and Environmental Sciences, University of Azores, Terceira Island, Portugal.
[f] Department of Agriculture and Food Science, Science and Research Branch, Islamic Azad University, Tehran, Iran.

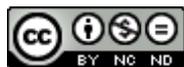

**Abstract:** All parts of the hawthorn tree (*Crataegus* spp*.*), including fruits, flowers, and leaves, have been used as a source of bioactive compounds. Thus, in this investigation, the twigs of two species of hawthorn plant of *Crataegus azarolus* (*C. azarolus*) and *Crataegus monogyna* (*C. monogyna*) were evaluated for bioactive compositions and biological activity (antioxidant and antimicrobial activities). To evaluate bioactive compositions, high-performance liquid chromatography (HPLC) was applied, and for biological activity, biochemical assays were performed. *C. monogyna* revealed a higher amount of total phenolic, total flavonoid, and total tannin contents compared to *C. azarolus.* The HPLC results indicated the highest amount of kaempferol (14.40%), catechin (17.70%), and gallic acid (25%) in twigs of *C. azarolus*, while the maximum quercetin (72%) compound was present in *C. monogyna*. *C. monogyna* exhibited higher antioxidant activity by 1,1-dizarophenyl-2-picrylhydrazyl (DPPH) (86.13%) and 2,2′-azino-bis (3-ethylbenzothiazoline-6-sulfonic acid (ABTS) (92.93%) compared to *C. azarolus* for antioxidant activity-DPPH (81.86%) and -ABTS (87.47%) assay. In the case of antimicrobial activity, the twigs of both species (especially *C. azarolus*) have a capacity against *Bacillus subtilis*, *Staphylococcus aureus*, and methicillin-resistant *Staphylococcus aureus*. The results of this study revealed that the twigs of both species contained a high amount of phenolic metabolites and antioxidant activity, while they showed low antimicrobial activity.

## 1. Introduction

Hawthorn trees, or *Crataegus* species, can be found all over the world, from Europe to Asia to North Africa to North America, and even northern Iraq [1]. Since ancient times, the plant's leaves, flowers, and fruits have all been employed in the practice of traditional medicine [2]. A lot of bioactive substances, like polyphenols, terpenoids, flavonoids, vitamins, and organic acids, are found in hawthorn fruit, leaves, and flowers [3]. These components enhance hawthorn's potential health advantages, including its anti-inflammatory, antioxidant, and immune-modulating properties, along with its capacity to support overall health [4, 5]. Several studies have highlighted the health benefits and antioxidant capacity of hawthorn fruits. In the early stages of liver and heart diseases, hawthorn is used to treat angina, high blood pressure, congestive heart failure, and cardiac arrhythmia. It is also a safe and effective treatment [6]. In addition to protecting cells from damage and promoting heart health, the high polyphenol content of hawthorn fruits is known to boost antioxidant activity. The presence of these components also results in the reduction of inflammation, which in turn lowers the risk of a variety of chronic diseases [3]. Due to its rich phytochemical profile and possible health advantages, hawthorn has recently attracted the attention of scholars [7-9].

Oxidative stress is a primary component contributing to chronic inflammation [10, 11]. This makes it easier for a number of transcription factors to become active, which then causes genes involved in inflammatory processes to be expressed [12]. Because of the kidneys' critical function in eliminating metabolic waste and maintaining chemical homeostasis, oxidative stress has the ability to cause various types of nephrotoxicity and renal impairment, which can have fatal consequences [13, 14]. An important strategy for preventing the oxidative stress is to consume foods that are high in antioxidants [15-17]. A potential preventive treatment that has been suggested is the addition of antioxidants such as vitamin E, vitamin C, tannins, phenolic acids, and flavonoids to meals [18]. This is because they can have anti-inflammatory, anti-carcinogenic, and anti-atherosclerotic properties when combined with other reducing agents [19]. In addition to their antioxidant and anti-inflammatory properties, polyphenols can block the enzymes responsible for eicosanoid production [20].

Supplemental antioxidants and plant extracts have shown promise in eliminating oxidative stress and inflammation caused by enzymatic and non-enzymatic components; some of these components are unique to plants and can only be consumed by the human body through food [21]. Moreover, foods that have been supplemented with bioactive compounds are able to enhance the immune system [22].

Limited evidence exists regarding the impact of integrating extracts from both hawthorn species of *Crataegus monogyna* (*C. monogyna*) and *Crataegus azarolus* (*C. azarolus*) on broad-spectrum bacteria and their bioactive chemicals.

In light of the information mentioned above, the present study attempted to evaluate the antimicrobial activity of both hawthorn species against a diverse panel of bacterial strains including *Bacillus subtilis* (*B. subtilis*), methicillin-resistant *Staphylococcus aureus* (*S. aureus*), and *S. aureus*. It also sought to assess the analysis of the metabolic profile of hawthorn through multiple assays including high-performance liquid chromatography (HPLC), free radical scavenging assays 1,1-dizarophenyl-2-picrylhydrazyl (DPPH) and 2,2′-azino-bis (3-ethylbenzothiazoline-6-sulfonic acid) (ABTS), total phenolic content, total flavonoid content, and total tannin content to provide a comprehensive profile of its antioxidant and anti-bacterial mechanisms. The results from this research may have implications for its potential applications in drug discovery as a functional dietary ingredient and as an alternative to conventional antibiotics, thereby contributing to the broader field of pharmaceutical industry research and development.

## 2. Materials and Methods
### 2.1. Samples Collection and Preparation

A total of 120 sets of twigs were obtained from *C. monogyna* and *C. azarolus* in the Sharbazher region of northern Iraq on July 25, 2024. Scientific recognition was conducted by Dr. Hoshman Omar Majeed (a taxonomist at Sulaimani University, College of Agricultural Engineering Sciences, Sulaimani, Iraq). The voucher specimens were then inserted into the herbarium of the institution. Upon collecting, the twigs were dried for 13 days at 25°C and preserved under dry and cool conditions before analysis. The experimental mill was used for the powdering of dry samples.

### 2.2. Sample Extraction by Maceration

The extracts were processed using a maceration technique. Forty-eight grams of powdered twigs per species were blended with 900 ml of solvent (a combination of 45% acetone, 45% methanol, and 10% dH2O) at 24°C for 20 h. Next, the extracts were centrifuged at 4000 rpm for 20 min, filtered, and vaporized using an evaporator at 40°C, then weighed and stored at 4°C.

### 2.3. Chemical Materials

Folin-Ciocalteu reagent, aluminum chloride, vanillin, DPPH, ABTS, and phenolic and flavonoid compounds were purchased from Sigma.

### 2.4. Bioactive Composition and HPLC Analysis

In relation to previous studies by Tahir *et al.* [23] and Lateef *et al.* [24], total phenolic (TPC) and total flavonoids (TFC) and total condensed tannin content (TTC) were calculated based on Folin-Ciocalteu, $AlCl_3$ and vanillin methods, respectively. Data are reported as gallic acid equivalents (mg GAE/g dry extract) for TPC, quercetin equivalent (mg QE/g dry extract) for TFC, and catechin equivalent (mg CE/g dry extract) for TTC. The sample was passed through a Whatman Nylon Membrane Filter (0.45 μm) and analyzed by HPLC (SYKAM-Germany). Pump model: S 2100 Quaternary Gradient Pump, Autosampler model: S 5200; Detector: UV (S 2340); Column Oven model: S 4115. The mobile phase was = (methanol: D.W: acetic acid) (85:13:2), the column was C18-ODS (25 cm × 4.6 mm), and detector UV–360 nm at a flow rate of 1mL/min.

### 2.5. Biological Activities of Twig Extracts
#### 2.5.1. Antiradical activity

The antioxidant capacity of the extracts was determined using DPPH and ABTS assays as described by Lateef *et al.* [24] and Tahir *et al.* [25]. In the determination of antioxidant activity by both methods, the extraction was prepared by mixing 0.1 g of hawthorn twig powder with 1 mL of 80% methanol. The mixture was shaken well for 30 minutes, and then incubated overnight at 5°C. The mixture was centrifuged at 10000 rpm for 10 minutes, and the supernatant was collected and stored at 5°C. In the first method, 150 μL of extract was mixed with 2 mL of DPPH solution (0.0038g DPPH dissolved in 100 mL of %95 methanol). The sample mixtures were incubated in the dark for 30 minutes at room temperature, with absorbance of the samples measured at 517 nm against the blank (95% methanol). While in the second method, 200 μL of extract was mixed with 2 mL 2,2ABTS [100 μL of the ABTS mixture diluted with 4.9 mL of 95% ethanol (1:50)]. The sample mixtures were incubated in the dark for 7 minutes at room temperature, with absorbance of the samples measured at 734 nm against the blank (95% methanol). A UV-visible spectrophotometer (UVM6100, MAANLAB AB, Sweden) was used in both cases.

#### 2.5.2. Antibacterial Capacity

The plant extract residues (100 mg) were dissolved in 2 mL of 2.5% dimethyl sulfoxide to achieve a final concentration of 0.05 mg/μL. Antibacterial tests were conducted using an agar well diffusion assay against certain bacterial strains as described by Tahir *et al.* [26]. Muller-Hinton agar was seeded with 100 μL of 0.5 McFarl and bacterial strains including *B. subtilis* (ATCC® 6633™), *S. aureus* (ATCC® 6538P™), and methicillin-resistant *S. aureus* (MRSA: Clinical isolates). Four wells were aseptically formed with a cork borer and each well was filled with 100 μL of the extract and incubated at 37°C for 24 h. After incubation, the inhibition zone (mm) was determined from the margin of the inhibition zone to the edge of the wells.

### 2.6. Statistical Analysis

One-way variance analysis (ANOVA) was conducted based on the results from each experiment using the program XLSTAT version 2019.2.2, accompanied by Duncan's new multiple range test at $p < 0.05$.

## 3. Results

### 3.1. Preliminary Phytochemical Screening

By using three chemical compounds as metrics, Table 1 compared the total phenolic content, total flavonoid content, and total tannin content for the two species (*C. azarolus* and *C. monogyna*). The values are presented in mg of compound per gram of dry weight (mg/g DW), and statistical differences between species are indicated. Regarding total phenolic content, *C. monogyna* exhibited higher phenolic content (22.23 mg GA/g DW) compared to *C. azarolus* (19.50 mg GA/g DW). This variation was statistically significant, as indicated by different superscripts. Similarly, *C. monogyna* showed higher total flavonoid content (1.23 mg QU/g DW) than *C. azarolus* (0.99 mg QU/g DW). For the last compound in the mentioned table, total tannin content, *C. monogyna* displayed a higher value of 6.98 mg CA/g DW compared to the value for *C. azarolus* of 6.29 mg CA/g DW.

**Table 1:** Total phenolic, flavonoid and tannin contents of twig extracts of the two plant species. Means that did not share a specific single letter above bars for each test were substantially different at P < 0.05 according to the Duncan test.

| Species | Total phenolic content (mg GA/g DW) | Total flavonoid content (mg QU/g DW) | Total tannin content (mg CA/g DW) |
|---|---|---|---|
| C. azarolus | 19.50 b | 0.99 b | 6.29 b |
| C. monogyna | 22.23 a | 1.23 a | 6.98 a |

### 3.2. Determination of Polyphenols Profile by HPLC

The HPLC results from figure 1 and table 2 showed that the two species were very different from each other. While *C. azarolus* exhibited a more diverse and evenly distributed chemical profile, *C. monogyna* was characterized by a high dominance of quercetin and a lack of kaempferol. The chemotaxonomic classification of the species or their use in targeted applications, such as pharmaceutical development or antioxidant studies, can be supported by these differences. In general, *C. azarolus* has a wider range of chemicals that can be found, such as Unknown, kaempferol, catechin, quercetin, and gallic acid. Flavonoids (kaempferol and quercetin) and phenolic chemicals (catechin, gallic acid) are present in large amounts in the mentioned species,. On the other hand, *C. monogyna* displayed fewer detectable compounds, with kaempferol being entirely absent and high abundance of quercetin and gallic acid, suggesting dominance of these compounds in its chemical profile. Retention time in our study indicates how long it takes for compounds to elute. In both studied species, it takes 2.59 min for a known compound. The compound's concentration and peak intensity were displayed by area and height, respectively, while area and height exhibited the relative contributions of compounds to the chromatogram. Regarding unknown substrates, *C. azarolus* has been shown to have higher area (589.75 mAU·s) and height (46.42 mAU) in comparison with *C. monogyna* (area: 249.28 mAU·s; height: 19.87 mAU). Similarly, the percentage contributions for *C. azarolus* were: 11.40% (area), 13.80% (height) although it was lower in C. monogyna: 6.10% (area), 7.10% (height). Kaempferol with retention time of 5.28 min was detectable in *C. azarolus* with area (746.60 mAU·s) and height (50.84 mAU) and was absent (not detected) in *C. monogyna*. Another compound of catechin (retention time: 6.51 min) was spotted in both species. *C. monogyna* had significantly lower area (252.34 mAU·s) and height (10.51 mAU) compared to *C. azarolus*: area (919.22 mAU·s) and height (42.23 mAU). This suggested *C. azarolus* had a much higher concentration of catechin. Significantly higher abundance of quercetin (retention time: 7.31 min) was found in *C. monogyna* with area (2965.65 mAU·s) and height (200.46 mAU) compared with *C. azarolus*, which had moderate area and height with values of 1641.11 mAU·s and 113.31 mAU, respectively.

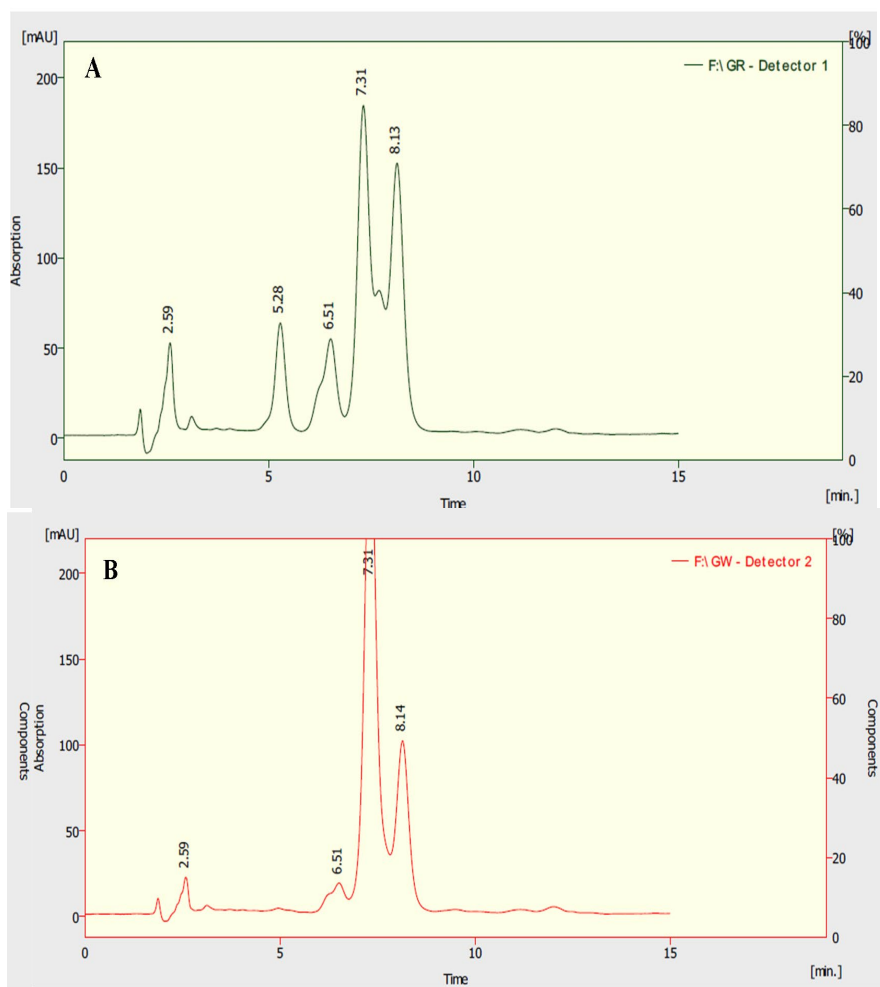

**Figure 1**: HPLC chromatogram of plant species. A: *C. azarolus*; B: *C. monogyna*.

**Table 2**: HPLC information on identifying compounds in plant species used in this investigation.

| Compound | Retention time (min) | Area (MaU.s) | Height (mAU) | Area (%) | Height (%) | W05 (min) |
|---|---|---|---|---|---|---|
| *C. azarolus* | | | | | | |
| Unknown | 2.59 | 589.75 | 46.42 | 11.40 | 13.80 | 0.20 |
| Kaempferol | 5.28 | 746.60 | 50.84 | 14.40 | 15.10 | 0.24 |
| Catechin | 6.51 | 919.22 | 42.23 | 17.70 | 12.60 | 0.31 |
| Quercetin | 7.31 | 1641.11 | 113.31 | 31.60 | 33.70 | 0.25 |
| Gallic acid | 8.13 | 1296.41 | 83.02 | 25.00 | 24.70 | 0.25 |
| *C. monogyna* | | | | | | |
| Unknown | 2.59 | 249.28 | 19.87 | 6.10 | 7.10 | 0.20 |
| Kaempferol | ND | ND | ND | ND | ND | ND |
| Catechin | 6.51 | 252.34 | 10.51 | 6.10 | 3.70 | 0.29 |
| Quercetin | 7.31 | 2965.65 | 200.46 | 72.10 | 71.50 | 0.25 |
| Gallic acid | 8.14 | 647.38 | 49.44 | 15.70 | 17.60 | 0.23 |

This work established the Duncan test as a way to identify particular differences among various sets of means concerning the four primary chemicals present in twig plant extract, as seen in figure 2. The three compounds, namely kaempferol, catechin, and gallic acid, evidently made significant contributions to *C. azarolus*. On the other hand, quercetin was the most often found component in C. *monogyna*.

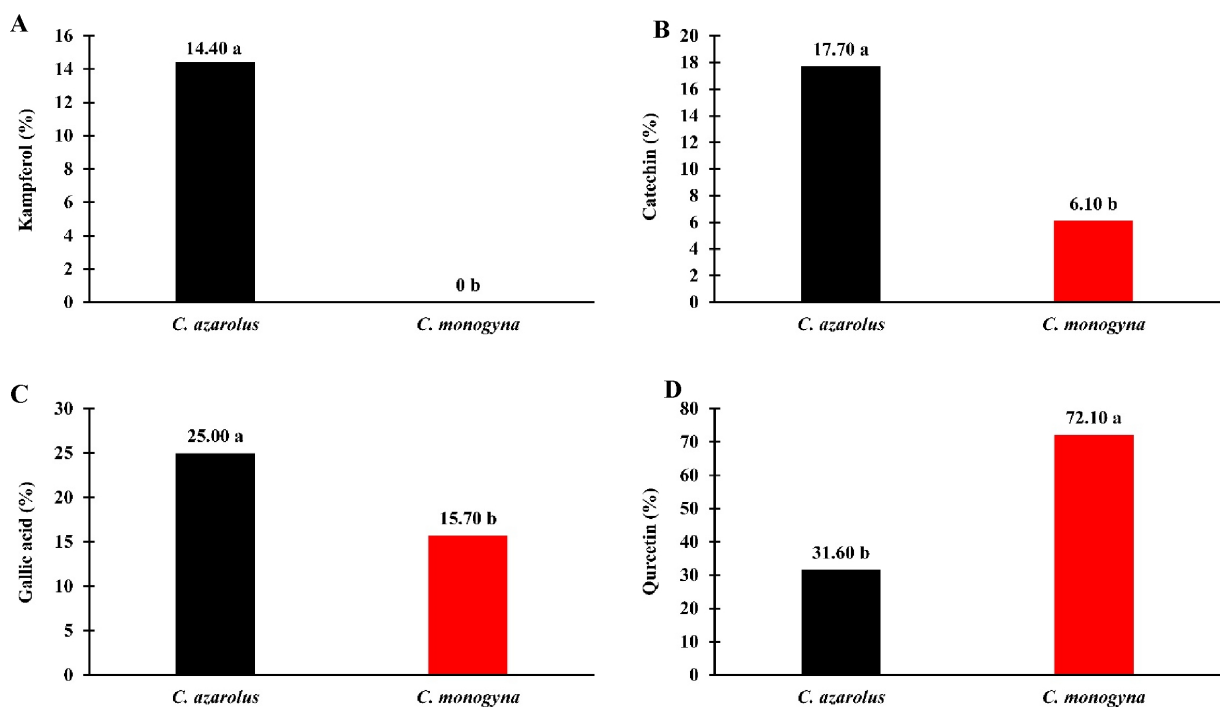

**Figure 2**: A: Kampferol, B: Catechin, C: Quercetin, and D: Gallic acid contents isolated from twig extracts of two plant species. Data are the average of triplicate measurements. Means that did not share a specific single letter above bars for each test were substantially different at p < 0.05 according to the Duncan test.

### 3.3. Biological Activity of Twig extracts

Table 3 indicated a comparative analysis of antiradical activities between the two species, *C. azarolus* and *C. monogyna*, measured using two assays: DPPH and ABTS (2,2'-azino-bis (3-ethylbenzothiazoline-6-sulfonic acid)) and the antibacterial activity of extracts from the two species, *C. azarolus* and *C. monogyna*, against three pathogenic bacterial isolates: *B. subtilis*, methicillin-resistant *S. aureus* (MRSA), and *S. aureus*. The values were specified as percentages, reflecting the antioxidant capacity of each species. The distinction between the two species was statistically significant for both DPPH and ABTS. *C. monogyna* exhibited higher antiradical activity (86.13%) compared to *C. azarolus* (81.86%) for DPPH assays. Similarly, *C. azarolus* showed less activity (87.47%) relative to *C. monogyna* (92.93%). The activity was measured by the diameter of the inhibition zone (in millimeters, mm), reflecting the efficiency of the extracts in inhibiting the tested bacterial growth. Regarding the antibacterial activity against *B. subtilis*, the inhibition zone for *C. azarolus* was 5.50 mm, slightly larger than that of *C. monogyna* with the value of 5.20 mm, while *C. azarolus* exhibited an inhibition zone of 5.83 mm, compared to 5.45 mm for *C. monogyna* in response to Methicillin-Resistant *S. aureus* (MRSA). A similar answer was noted against *S. aureus* by way of *C. azarolus*, again displaying slightly higher activity (6.56 mm) compared to *C. monogyna* (6.25 mm).

**Table 3**: Antiradical activities (DPPH and ABTS) of twig extracts and inhibition zone (mm) of four twig extracts against three human pathogenic bacteria. Data are the average of triplicate measurements. Means that did not share a specific single letter above bars for each test were significantly different at P < 0.05 according to the Duncan test.

| Species | Antioxidant activity-DPPH (%) | Antioxidant activity-ABTS (%) | against *B. subtilis* (mm) | methicillin-resistant *S. aureus* (mm) | *S. aureus* (mm) |
|---|---|---|---|---|---|
| *C. azarolus* | 81.86 b | 87.47 b | 5.50 a | 5.83 a | 6.56 a |
| *C. monogyna* | 86.13 a | 92.93 a | 5.20 a | 5.45 | 6.25 a |

### 4. Discussion

The highest amounts of total content of three categories: phenol, flavonoid, and tannin, were exhibited by the two species. The differences in the content between the three categories is due to the genetic makeup of the species. In general, *C. monogyna* constantly determined higher values across all three chemical categories, suggesting it may possess more robust antioxidant properties than *C. azarolus*. The outcome of HPLC indicated that quercetin is the dominant compound in *C. monogyna*. The final substrate, gallic acid, in both species required higher retention times, with the value of 8.14 in association with previously mentioned substrates. This evidence indicates gallic acid is more abundant in both studied species. The overall findings of this investigation were that *C. azarolus* shows a more sensible distribution of compounds, with significant contributions from quercetin, gallic acid, and catechin. Quercetin, on the other hand, in *C. monogyna* is dominated by over 70% contribution. Substrate superstation efficiencies (ranging between 0.20 and 0.31 min) showed that the two species have similar chromatographic resolution. The unique presence of kaempferol in *C. azarolus* suggests it could be a distinguishing compound for this species.

The antibacterial activities of both plant extract species were comparable across all three bacterial isolates, with *C. azarolus* representing slightly higher inhibition zones. However, the differences are not statistically significant as both shared the same superscripts. The phenolic, flavonoid, and tannin content in the extracts likely contributes to their antimicrobial properties. Higher levels of these compounds in *C. monogyna* may suggest potential variability in antibacterial mechanisms. Differences in cell wall structures and resistance mechanisms among the isolates (e.g., MRSA's resistance to methicillin) could influence susceptibility to the extracts. The combined action of multiple bioactive compounds in the extracts may result in a more pronounced or similar antibacterial effect. The results suggest the efficacy of *C. azarolus* and *C. monogyna* as antioxidant and antibacterial agents, particularly for addressing human illnesses. Further investigation into specific active compounds, extract concentrations, and their modes of action is essential to enhance and comprehend their antioxidant and antibacterial efficacy. The antioxidant and antibacterial efficacy of plant extracts, including those from both utilized plants, is generally facilitated by bioactive chemicals that affect bacterial structures or interfere with essential metabolic pathways [27]. Phenolics have been demonstrated to damage cell membrane integrity, resulting in bacterial mortality owing to osmotic imbalance [28]. Phenolic compounds, such as flavonoids and tannins, have the ability to disrupt the bacterial cell wall or membrane through their interactions with proteins or lipid bilayers [29]. This results in heightened permeability and the leakage of cellular content. A good example would be the binding to peptidoglycan, which weakens the cell walls of bacteria [30]; an integration into lipid bilayers, which results in the destabilization of membranes and the lysis of cells [31]. Flavonoids such as quercetin are recognized for their ability to inhibit bacterial DNA gyrase, a crucial enzyme for replication and transcription [32]. Flavonoids and tannins have the potential to bind to bacterial enzymes, thereby disrupting critically important metabolic processes, including the replication of DNA, ATP production, and the synthesis of proteins [33].

Overall, *C. monogyna* indicated greater antiradical activity in both the DPPH and ABTS assays, suggesting that it possesses stronger free-radical scavenging abilities compared to *C. azarolus*. This lines up with its higher bioactive compound content as mentioned in table 2. Under specific circumstances, antioxidants from phenolic-rich plants can function as pro-oxidants, resulting in the generation of reactive oxygen species (ROS) and the breakdown of bacterial cells [34]. Extracts that are high in antioxidants kill microbes by causing oxidative stress in bacterial cells [35]. This is done by making reactive oxygen species (ROS), which damage bacterial DNA, proteins, and lipids. This reactive stress is too much for the bacteria's defenses to handle, and the cells eventually die [36]. In *S. aureus* strains that are resistant to multiple drugs, tannins have been reported to inhibit efflux pumps, which results in an increase in the effectiveness of antibacterial agents [37].

Research has demonstrated that flavonoids reduce the pathogenicity of *Pseudomonas aeruginosa* by inhibiting quorum sensing. Quorum sensing is a communicative process employed by bacteria to modulate collective behaviors, including biofilm development and the synthesis of virulence factors [38]. There is some evidence that phenolics and flavonoids can inhibit bacterial pathogenicity and biofilm formation by interfering with quorum sensing molecules, such as acyl-homoserine lactones [39, 40]. In our analysis, there is a great chance that the presence of these chemical substances was closely associated with these pathways, which resulted in the inhibition of the colonization of the bacterial isolates that were brought under investigation. This is because these pathways led to the inhibition of the colonization.

Both species show promising antimicrobial potential, but there are notable limitations that affect the applicability and reliability of these finding. Most studies are in vitro (lab-based), testing extracts against bacteria in petri dishes. These results often do not translate directly to clinical efficacy in the human body, where factors like metabolism, absorption, and immune response come into play.
It is well-documented that hawthorn species include tannins, organic acids, and flavonoids that have antibacterial and anti-inflammatory properties [2]. At the other end of the spectrum, the polyphenols and flavonoids found in Hawthorn species have been shown to be effective against Gram-positive and Gram-negative bacteria [4].

In order to get important insights into the possible applications of these plants in the treatment of bacterial infections, additional studies into combination applications, such as co-formulation for infection therapy or comparative examinations of these plants, are required.

## 5. Conclusions

The findings of this study provide proof that the *C. azarolus* and *C. monogyna* hawthorn species possess antibacterial and antioxidant capabilities. Broad-spectrum antibacterial activity was established by the plant extracts from both species against the three bacterial strains that were examined, which were *B. subtilis*, methicillin-resistant *S. aureus*, and *S. aureus*. Additionally, they demonstrated considerable antioxidant activity, which enabled them to successfully scavenge radicals such as hydroxyl, superoxide anion, and DPPH compounds. Significant amounts of total phenolic content, total flavonoid content, and total tannin content were found in both plant extract species, with *C. monogyna* exhibiting the highest amounts in terms of the three substrates mentioned earlier. The chemical profile of both plant extracts revealed a significant presence of four substrates: kaempferol, catechin, quercetin, and gallic acid. According to these findings, the extracts of the hawthorn's plant species in this study may have significant applications as a source of natural antioxidants and as an agent against a wide variety of bacterial pathogenicity.
Isolating specific bioactive chemicals to determine their exact processes should be the primary goal of future research; conducting tests against a wider variety of diseases to evaluate the effectiveness of the spectrum.

**Author contribution**: **Hiwa Sheikh Ahmed**: Data curation, Investigation, Methodology; **Kadhm Abdullah Muhammad**: Data curation, Investigation, Methodology; **Djshwar Dhahir Lateef**: Data curation, Investigation, Methodology, Resources; **Kamaran Salh Rasul**: Data curation, Investigation, Methodology, Project administration; **Abdulrahman Smail Ibrahim**: Software, Validation, Visualization, Writing – original draft; **Mariana Casari Parreira**: Software, Validation, Visualization, Writing – original draft; **Weria Weisany**: Supervision, Writing – original draft, Writing – review & editing

**Data availability**: Data will be available upon reasonable request.

**Conflicts of interest**: The authors declare that they have no known competing financial interests or personal relationships that could have appeared to influence the work reported in this paper.

**Funding**: The authors did not receive support from any organization for the conducting of the study.